\documentclass[preprint]{aastex}
\slugcomment{To Appear in: The Astrophysical Journal}
\shortauthors{Nemiroff}
\shorttitle{GRB Pulse Scale Conjecture}

\begin{document}

\title{The Pulse Scale Conjecture and the Case of BATSE Trigger 
2193}

\author{Robert J. Nemiroff} 

\affil{Michigan Technological University, Department of Physics, 1400 
Townsend Drive, Houghton, MI 49931}
\email{nemiroff@mtu.edu}

\begin{abstract}
The pulses that compose gamma-ray bursts (GRBs) are hypothesized to have the same shape at all energies, differing only by scale factors in time and amplitude.  This ``Pulse Scale Conjecture" is confirmed here between energy channels of the dominant pulse in GRB 930214c (BATSE trigger 2193), the single most fluent single-pulsed GRB that occurred before May 1998. Furthermore, pulses are hypothesized to start at the same time independent of energy.  This ``Pulse Start Conjecture" is also confirmed in GRB 930214c.  Analysis of GRB 930214c also shows that, in general, higher energy channels show shorter temporal scale factors.  Over the energy range 100 KeV - 1 MeV, it is found that the temporal scale factors between a pulse measured at different energies are related to that energy by a power law, possibly indicating a simple relativistic mechanism is at work.  To test robustness, the Pulse Start and Pulse Scale Conjectures were also tested on the four next most fluent single-pulse GRBs.  Three of the four clearly passed, with a second smaller pulse possibly confounding the discrepant test.  Models where the pulse rise and decay are created by different phenomena do not typically predict pulses that satisfy both the Pulse Start Conjecture and the Pulse Scale Conjecture, unless both processes are seen to undergo common time dilation. 
\end{abstract}


\section{Introduction}

The physical mechanisms that create gamma ray bursts (GRBs) and their constituent pulses remain unknown even 30 years after their discovery.  Furthermore, GRBs may make powerful tools uniquely visible into 
the early universe, particularly if it is possible to accurately estimate their 
distances.   Recent redshift determinations of the optical counterparts to 
GRBs have placed them at cosmological distances \citep{Djo97,Met97}, 
but remain scarce, currently numbering on order ten.  \citet{Nor00} have 
suggested that GRB pulses might calibrate GRB intrinsic luminosity, 
allowing perhaps a few hundred of the brightest GRBs currently detected by the Burst and Transient Source Experiment (BATSE) on board the Compton Gamma-Ray 
Observatory (CGRO) to act as standard candles.  Specifically, \citet{Nor00} have 
proposed that the time-lag between BATSE DISCSC energy channels 3 and 1, as 
measured by a cross-correlation function, might calibrate this
standard candle.  More recently, \citet{Fen00} has proposed that 
general variability across the whole GRB may also act to calibrate these 
explosions as standard candles.  The data behind these claims remains small, however, and proposed physical mechanisms behind them unproven.

\citet{Des81} originally noted that GRBs have characteristic ``separated times" in events, an illusion to structures that are today known as pulses.  \citet{Nor96} recently fit and analyzed 
the pulses in 41 BATSE GRBs with a minimal functional form.  
\citet{Kat94} has suggested that the shape of a GRB pulses originates from 
time delays inherent in the geometry of its spherically expanding emission 
front. \citet{Lia97} have provided arguments that saturated Compton 
up-scattering of softer photons may be the dominant physical mechanism that 
creates the shape of GRB pulses.  \citet{Sum98,Ram99,Fen00} claim that 
the rise time of GRB Fast Rise Exponential Decay (FRED)-like structures is 
related to the sound speed of the pulse medium, but the decay time is related 
to a time-delay inherent in the geometry of an expanding, spherical GRB 
wave front undergoing rapid synchrotron cooling.

How the spectrum of GRBs change with time also has a long history 
\citep{Whe73,Ved81,Nor83,Cli84,Nor86}.  Recent notable work includes \citet{Ban97} who used auto- and cross-correlation statistics to track hardness variability in 209 bright BATSE bursts, and \citet{Cri99} who tracked the peak of photon frequency times flux of 41 pulses in 26 bursts, finding that this quantity decays linearly with pulse energy fluence.

In this paper a bright GRB dominated by a single smooth pulse is analyzed 
in an effort to better understand GRB pulses in general, and more specifically to help calibrate energy dependent pulse attributes that are being used as standard candles for GRBs. These attributes may also help distinguish or eliminate some physical processes proposed to create GRB pulses.  In \S 2 data from BATSE trigger 2193 will be introduced and discussed.  In \S 3 the Pulse Start Conjecture will be proposed.  In \S 4, the Pulse Scale Conjecture will be introduced, discussed, and tested on data from 2193.  In \S 5 the next four most fluent GRBs will also used to test the Pulse Start and Pulse Scale Conjectures.  In \S 6, the paper will conclude with estimates of the theoretical implications following from the potential truths of these conjectures.

\section{A Case Study: GRB 930214c -- BATSE Trigger 2193}

This project started with the search for a single, isolated, fluent, long GRB 
pulse to study in detail.  The single pulse that appears to compose GRB 930214c (BATSE trigger 2193) was chosen because of its isolated nature, high fluence, and long duration.  A list of GRBs with durations greater than two seconds that are dominated by a single pulse has been published by \citet{Nor99} as Table 1.  BATSE trigger 2193 has the highest fluence on this table, as computed by multiplying the Full Width Half Maximum of the duration (column 4) by peak counts $A_{max}$ (column 6).  The isolated nature of the main pulse in 2193 minimizes confusion by simultaneously arriving pulses.  The relatively high fluence of this pulse makes for relatively good statistics.  The relatively long duration of this pulse makes the time lag between BATSE energy channels proportionally longer.  The combined fluence and duration of GRB 930214c allow BATSE CONTINUOUS (CONT) data to be useful interesting time and energy scales are resolved.  CONT data is divided into 16 energy channels (as opposed to only 4 energy channels for the more frequently analyzed DISCLA and DISCSC data types) but only time-resolved into 2.048 second longs time bins.  Important data from before BATSE triggered is also available with the CONT data type.  Additionally, CONT data is available in background subtracted form from the Compton Gamma 
Ray Observatory Science Support Center web pages.  Specifically, 
background subtracted CONT data for 2193 created by G. Marani 
is available from {\url http://cossc.gsfc.nasa.gov/cossc/batse/}.

Light curves for BATSE trigger 2193 are shown for the four main broad-band 
energy channels of PREB, DISCSC, and DISCLA data in Figure 1, and for 
all 16-energy channels of CONT in Figure 2.  The four energy channels of DISCSC data and displayed in Figure 1 are, approximately: channel 1: 25-60 KeV, channel 2: 60-110 KeV, channel 3: 110-325 KeV, and channel 4: 325 KeV - 1 MeV.  These data have relatively course energy resolution but with 64-ms time bins, relatively high time resolution.  The sixteen CONT energy channels in Figure 2 are: channel 1: 13-25 KeV, channel 2: 25-33 KeV, channel 3: 33-41 KeV, channel 4: 41-54 KeV, channel 5: 54-72 KeV, channel 6: 72-96 KeV, channel 7: 96-122 KeV, channel 8: 122-162 KeV, channel 9: 162-230 KeV, channel 10: 230-315 KeV, channel 11: 315-424 KeV, channel 12: 424-589 KeV, channel 13: 589-745 KeV, channel 14: 745-1107 KeV, channel 15: 1107-1843 KeV, and channel 16: 1843-100,000 KeV.  These energy boundaries are only approximate, as each of BATSE's eight detectors has slightly different energy boundaries.

Inspection of Figures 1 and 2 show that the counts are highest in the 
middle range energies detectable by BATSE.  The highest fluence is in CONT energy discriminator channel 9.  It is clear from both plots that this pulse has the classic Fast Rise Exponential Decay (FRED) shape that is common to many 
pulses in GRBs.

\section{The Pulse Start Conjecture}

Inspection of Figure 2 shows that the pulse begins at approximately the same time in all 16 CONT energy channels, with an approximate error of one 2.048-second duration time bin. The simultaneous start time is in marked contrast to the time of peak flux in each energy channel, which clearly changes as a function of energy.  It is here hypothesized that this behavior is common to all GRB pulses, and it will be referred to here as the ``Pulse Start Conjecture." 

The Pulse Start Conjecture is something that may have been obvious to 
some previously but not discussed explicitly.  Its truth, however, may be 
important to accurate pulse modeling paradigms.  The Pulse Start Conjecture is also important to proper application of the following Pulse {\it Scale} Conjecture, since it identifies the zero point that is held fixed when scaling the time axis.  Note that this start time is generally {\it not} the same time as the BATSE trigger time, which is an instrument dependent phenomenon. In general, if a pulse triggers BATSE, its start time will usually occur before the BATSE trigger.  If the pulse did not trigger BATSE, its start time will usually occur after the BATSE trigger.

\section{The Pulse Scale Conjecture}

It appears from an informal inspection of Figures 1 and 2 that pulse shapes 
in each energy channel are similar.  In this section it is 
tested whether the shape in each discriminator energy channel is 
actually {\it statistically} similar to the shape is all other channels, given  single scale factors in time and amplitude.  In other words, can any light-curve strip in Figure 2 at any energy be stretched along both the x-axis (time) and the y-axis (amplitude) to precisely overlay any other light-curve strip at any other energy?  When stretching in the time direction, the zero point (which is fixed) is the energy-independent start time of the pulse.  When stretching in the counts direction (amplitude), the zero point is the background level of each burst.

To test the Pulse Scale Conjecture, the light curves of CONT energy channels 4 through 14 of GRB 930214c were stretched to their best-fit match to CONT energy channel 9, the channel with the highest peak count rate.  Twenty-six 2.048-sec CONT time bins were isolated from for each CONT energy channel, noting the fit background counts and GRB counts for each time bin.  The time series was constructed so that the first bin was the start of each pulse, regardless of energy, in accord with the Pulse Start Conjecture.  This zero point was held fixed during all computational time-contractions.  Channel 9 was isolated, and compared to other candidate energy channels.  A candidate temporal scale factor was iterated between channel 9 and this energy channel.  The time bins available for statistical comparison were identified.  Lower energy channels were then artificially time-contracted and re-binned to the candidate scale of channel 9, and normalized by the available fluence.  A $\chi^2$ statistic per degree of freedom was then computed and recorded.  All energy channels were iterated, as were 100 candidate scale factors.  Higher energy channels were compared similarly with the exception that channel 9 was artificially time-contracted to the higher-energy scale.

The results are shown in Figure 3.  Data points from each energy channel are plotted with the channel number, while a solid line depicts channel 9.  When the best-fit dilation factors in time and amplitude are applied, the 
pulse shapes match that of channel 9 with the following $\chi^2$ per degree 
of freedom: (0.82, 2.86, 8.60, 1.80, 1.26, 1.71, 0.73, 1.02, 1, 1.08, 2.43, 0.85, 1.63, 1.17, 4.46, N/A).  The last energy channel, 16, had so few counts that accurate results could not be obtained.  Channels 2, 3, and 15 also had relatively few counts (as visible in Figure 2), which could create an underestimated variance which in turn could create an overestimated $\chi^2$ when compared to channel 9.  Most of the energy channels, however, give a relatively good fit, indicating that the pulse shape in most energy channels really is statistically similar to the pulse shape in channel 9.  Most energy channels therefore appear to adhere to the Pulse Scale Conjecture. 

Figure 4 shows a plot of the scale factor in time needed to best bring a 
pulse in a given energy channel into alignment with discriminator channel 9, 
plotted as a function of the geometric mean energy of the energy channel.  The error bars in the y-direction indicate the range of scale factors below a $\chi^2$ per degree of freedom of 2.5.  The error bars in the x-direction denote the CONT energy channel boundaries. 

Inspection of Figure 4 yields several interesting features.  First, in general, the higher energy channels appear time-contracted relative to lower energy channels.  That pulses have shorter durations at higher energies has been noted previously (e.g. \citet{Nor86}). Such scaling behavior possibly indicates a relativistic origin for the relative apparent time contraction of the higher energy channels. 

Next, inspection of the mid-energy channels in Figure 4 reveals a nearly linear relation between the log of the scale factor in time and the log of the mean energy of counts. This power law dependence appears to hold from at least channel 7 through channel 14, from about 100 KeV to about 1000 KeV. This unexpected relationship may point to a relatively simple progenitor relation between photon energy and relativistic time dilation factor.

Figure 5 shows a plot of the scale factor in amplitude needed to best bring 
a pulse in a given energy channel into alignment with discriminator channel 
9, plotted as a function of the geometric mean energy of the energy channel.  
We note that the pulse appears to have a lesser amplitude in both higher and 
lower energy channels.  This is not surprising, as channel 9 was initially 
chosen just because it had the highest peak flux (and hence best counting 
statistics).  Figure 5 is essentially a normalized 16-channel spectrum of the counts received from this GRB.  No easily recognizable relation between the scale factors in time and energy has been found. 

\section{The Next Four Most Fluent Single-Pulse GRBs}

Possibly BATSE trigger 2193 is an unusual burst.  Perhaps the main pulse in 2193 is the only one to approximately fit the Pulse Start and Pulse Scale Conjectures.  To test uniqueness, the next four most fluent pulses in single-pulse GRBs as listed in \citet{Nor99} were analyzed.  These were, in order of decreasing fluence, GRB 930612a, GRB 940529b, GRB 941031b, and GRB 970825, corresponding to BATSE trigger numbers 2387, 3003, 3267, and 6346 respectively.  Because these GRBs have significantly fewer counts, it was worried that passing the sixteen-channel test would be significantly less demanding due to inherently larger variances.  It was therefore decided to use four energy channel data, also downloadable from the Compton Gamma Ray Observatory Science Support Center.  

Each GRB was prepared in a manner similar to that described above for BATSE trigger 2193.  Here, however, it was tested to see if the pulse shape in 
energy channels 1 and 2 could be scaled in time and fluence to match the pulse shape found in energy channel 3.  The number of counts was so few in energy channel 4 for each burst that any test involving data from this channel was essentially meaningless.  A $\chi^2$ per degree of freedom ($d$) statistic was computed between the two lower energy channels and channel 3.  For BATSE trigger 2387, the $\chi_{13}^2$ per degree of freedom between channels 1 and 3 was 5.06, while between channels 2 and 3 $\chi_{23}^2/d = 2.20$.  For BATSE trigger 3003, $\chi_{13}^2/d = 1.25$ and $\chi_{23}^2/d = 1.07$.  For BATSE trigger 3267, $\chi_{13}^2/d = 0.98$ and $\chi_{23}^2/d = 1.09$.  Lastly, for BATSE trigger 6346, $\chi_{13}^2/d = 0.72$ and $\chi_{23}^2/d = 0.74$.  These tests were all carried out using 64-ms time-binned data.

Plots where the data is binned to 1.024-second bins are shown in Figures 6-9. Trends in the data are easier for the human eye to discern on this large time-bin size.  The best-found scale factors between the first two energy channels and channel 3 has been applied.  Channel 3 data is plotted with a continuous line, while channel 1 and channel 2 data are plotted with the symbols ``1" and ``2" respectively. 

From these results, it appears that the combination Pulse Start and Pulse Scale Conjectures hold in all cases with the possible exception of BATSE trigger 2387.
Even for BATSE trigger 2387, the Pulse Start and Pulse Scale Conjectures were marginally consistent between energy channels 2 and 3.  It is speculated that the conjecture tests might have been compromised between channels 1 and 3 because of a second dim soft pulse that occurred well after the peak of the main pulse.  

\section{Summary, Theoretical Implications, and Conclusions}

From preliminary inspection of the most fluent single-pulse GRB on the \citet{Nor99} list, BATSE trigger 2193, two conjectures have been suggested.  
They are: 1.  The Pulse Start Conjecture: GRB pulses each have a unique 
starting time that is independent of energy. 2. The Pulse Scale Conjecture: GRB pulses have a unique shape that is independent of energy.  The relation between the shape of a GRB pulse at any energy and the shape of the same GRB pulse at any other energy are simple scale factors in time and amplitude.  

These two conjectures were then tested on BATSE trigger 2193 and found statistically valid in significantly fluent energy channels.  The two conjectures were also indicated as true for three of the next four most fluent single-pulse GRBs, with the discrepant case possibly being affected by the presence of a small secondary pulse.  When the Pulse Scale conjecture holds, a unique time-scaling factor is revealed between GRB energy bands.  For BATSE trigger 2193, this scale factor was found to be a power law increasing monotonically from about 100 KeV to 1000 KeV.  

One might be surprised that the Pulse Scale Conjecture can be tested at all with current BATSE data, since the energy channels available all have finite energy width.  In its purest form, the Pulse Scale Conjecture predicts a scaling relation between pulse shapes at monochromatic energies only.  In this sense, it is fortunate that the shape of GRB pulses allows pulses in different energy channels to be added together and nearly retain their initial shape.  Were this not true, the broad width of BATSE energy channels would have made such a behavior indiscernible.  Mathematically, however, the superposition does not work exactly, which may mean that the conjectures works even more precisely than indicated here. 

Much research has been done isolating GRB pulses and tracking them across energy channels \citep{Nor96, Sca98}.  The truth of the Pulse Scale conjecture in other GRB pulses might provide useful constraints on GRB pulse identification and energy tracking.

Several important works involving GRB pulses have been published which 
involve a cross correlation between pulses in different energy bands 
\citep{Nor00, Fen00}.  The accuracy of these works might be augmented 
were they to factor in a best-fit scale-factor before computing a cross-
correlation.  In the case of \citet{Nor00}, this might result in a more accurate 
standard candle for GRBs.

The Pulse Scale Conjecture identifies a potential invariant of each GRB 
pulse: the ratio of the fluence occurring before the peak to the fluence 
occurring after the peak.  It has long been known that GRBs are time-asymmetric \citep{Nem94}, but the energy-invariant degree of asymmetry of GRB pulses might now be used as a tool.  Since neither pulse scale factors in time nor in amplitude change this quantity, each independent GRB pulse should have its own invariant pre-/post- peak fluence ratio.  If different physical processes determine the duration of the rise and decay separately, then one would not expect that these processes would scale the same at different energies.  Therefore, the constancy of this rise/decay ratio over different energy bands indicates either that the same physical process creates the rise and decay, or that both processes are effected by a common dilation.

Beyond this, the Pulse Scale Conjecture may be interpreted as a relative 
time-dilation factor between energies.  One might speculate that this 
behavior results from a relative projected speed or a relative beaming 
angle, but these speculations cannot yet be definitively tested.

The Pulse Scale Conjecture also gives insight into why some pulses 
appear to have hardness decrease monotonically with time 
\citep{Whe73,Nor83,Lar85,Nor86,Cri99}, while some pulses appear to have hardness 
track intensity \citep{Ved81,Cli84,Gol83,Nor86,Cri99}.   In the former 
case, a relatively small temporal scale-factor might exist between the high and low energy channels.  A small temporal scale-factor is nearly equal to a constant shift to an earlier time of the higher energy band relative to the lower energy band.  For a FRED shaped pulse, such a shift will appear as monotonically decreasing hardness in coarsely binned timing data.  

Pulses where hardness appears to track intensity, in contrast, should exhibit a 
relatively large temporal scale-factor between hardness compared energies. A large scale-factor would allow the pulse at the higher energy to peak while the 
same pulse at lower energy is still rising slowly.  To an approximation, the 
pulse intensity at the lower energy is flat, and so the ratio of the high to low 
energy channels is nearly proportional to the intensity of the high-energy 
channel.

Future work might try to discern the truth of the Pulse Start and Pulse Scale conjectures over a wider variety of pulses.  No other GRB pulses were tested in detail and dismissed because they failed the Pulse Start and Pulse Scale Conjectures stated above.  It is true, however, that most GRB pulses occur simultaneously with other pulses, and so would be much more difficult to test.  Informal inspection of isolated pulses in several other GRBs does indicate that the conjectures proposed here may be applicable to many other -- perhaps all other -- GRB pulses. 

\acknowledgements

This research was supported by grants from NASA and the NSF.  I would 
like to thank Jay Norris, Jerry Bonnell, and Christ Ftaclas for many helpful 
discussions.

\clearpage

\figcaption[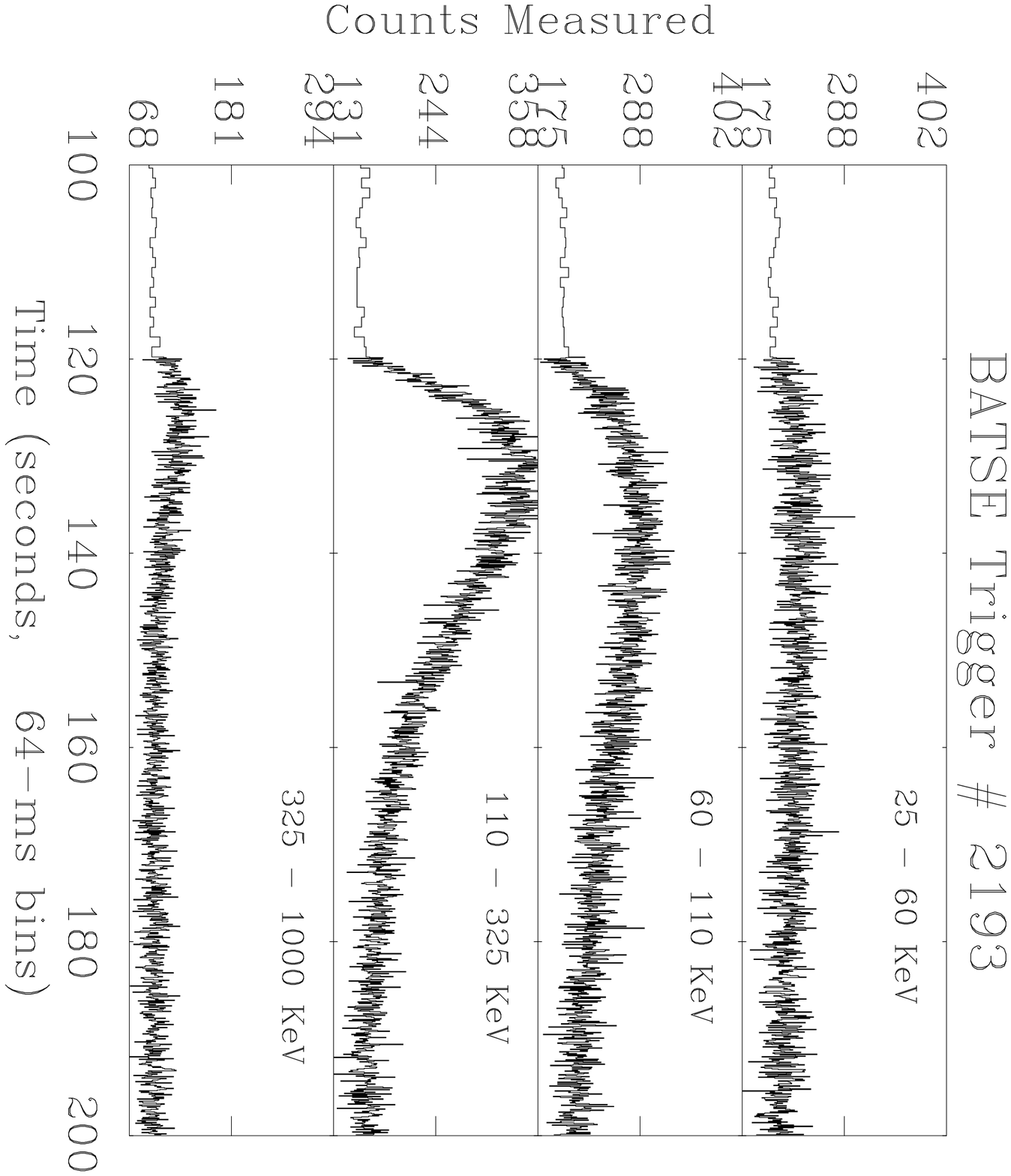]
{A light-curve plot of BATSE trigger 2193 in the four broad energy channels of PREB, DISCDC, and DISCLA data.  The zero point of time indicated here is arbitrary.  Trigger time occurs 2.048 seconds after the time resolution has switched from 1.024 seconds to 64-ms. 
\label{fig2}}

\figcaption[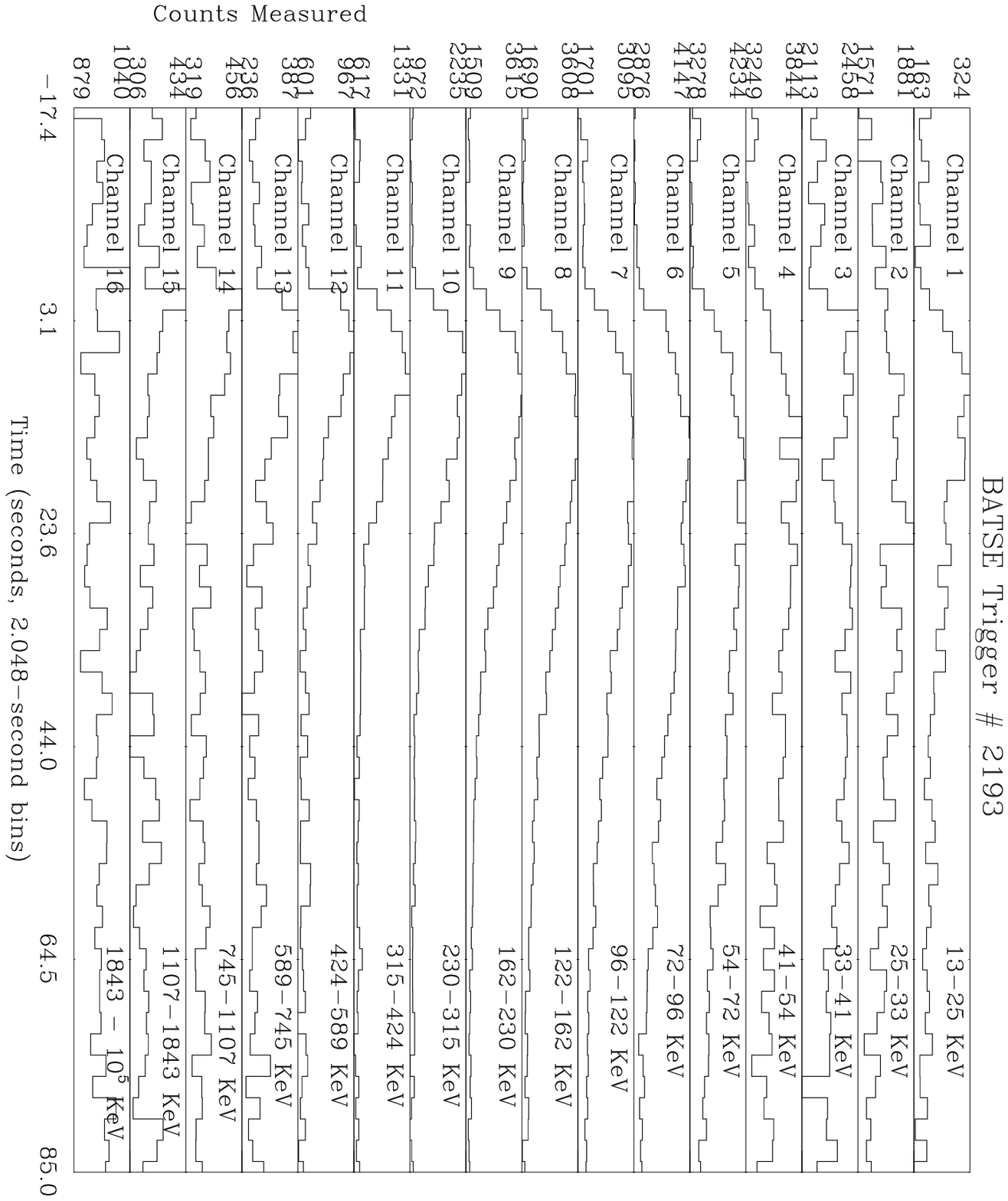]
{A light-curve plot of BATSE trigger 2193 in the sixteen CONT data channels.  Trigger time occurs when time equals zero.
\label{fig1}}

\figcaption[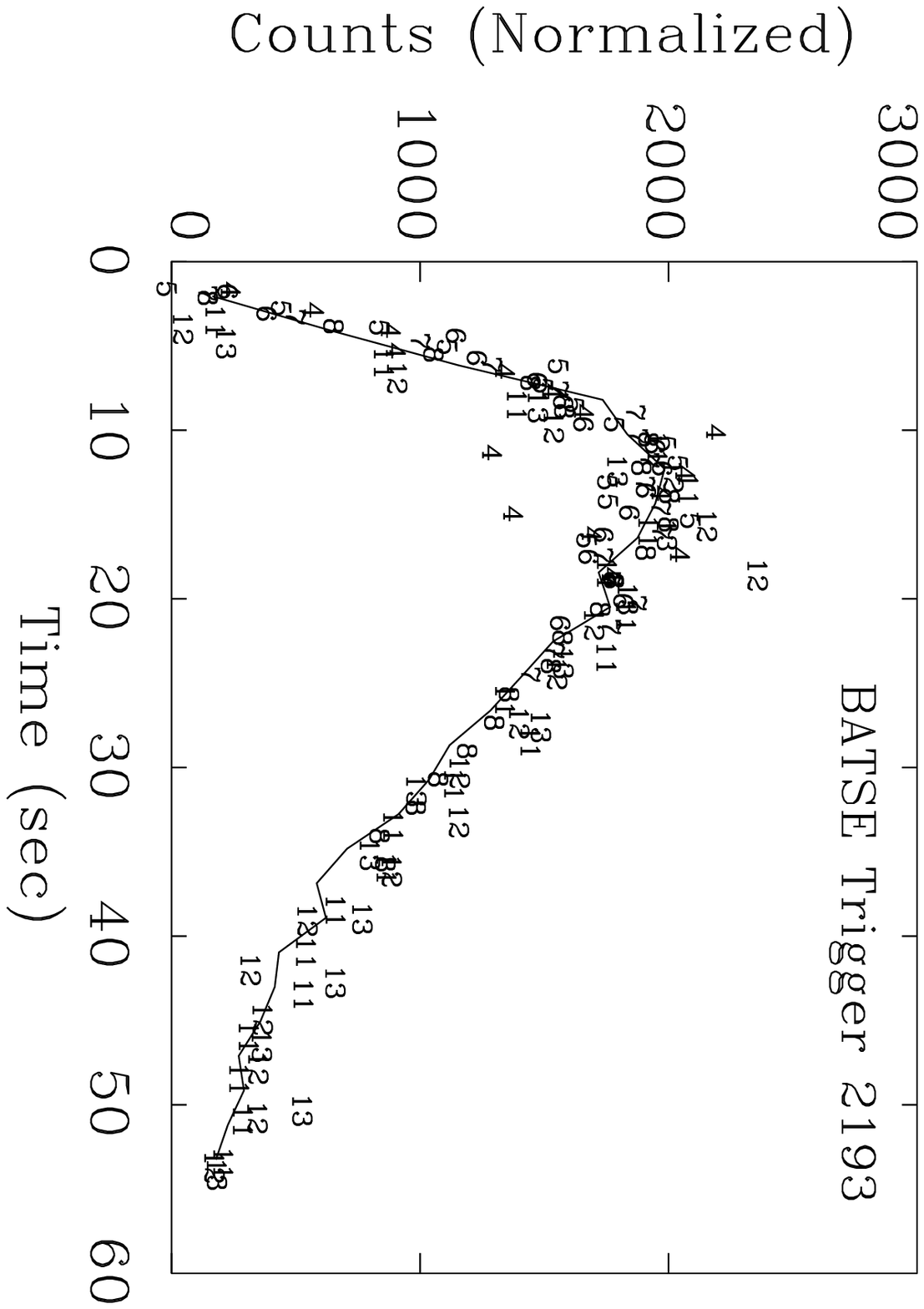]
{A light-curve plot of BATSE trigger 2193 with central energy channels scaled to BATSE discriminator channel 9.  BATSE channel 9 is drawn in as a 
continuous line. The scaled counts from each energy channel are plotted with numbers indicative of each CONT energy channel.  Time has been set to zero at the beginning of the pulse.  Note how the scaled pulse in each energy channel has the same shape.
\label{fig3}}

\figcaption[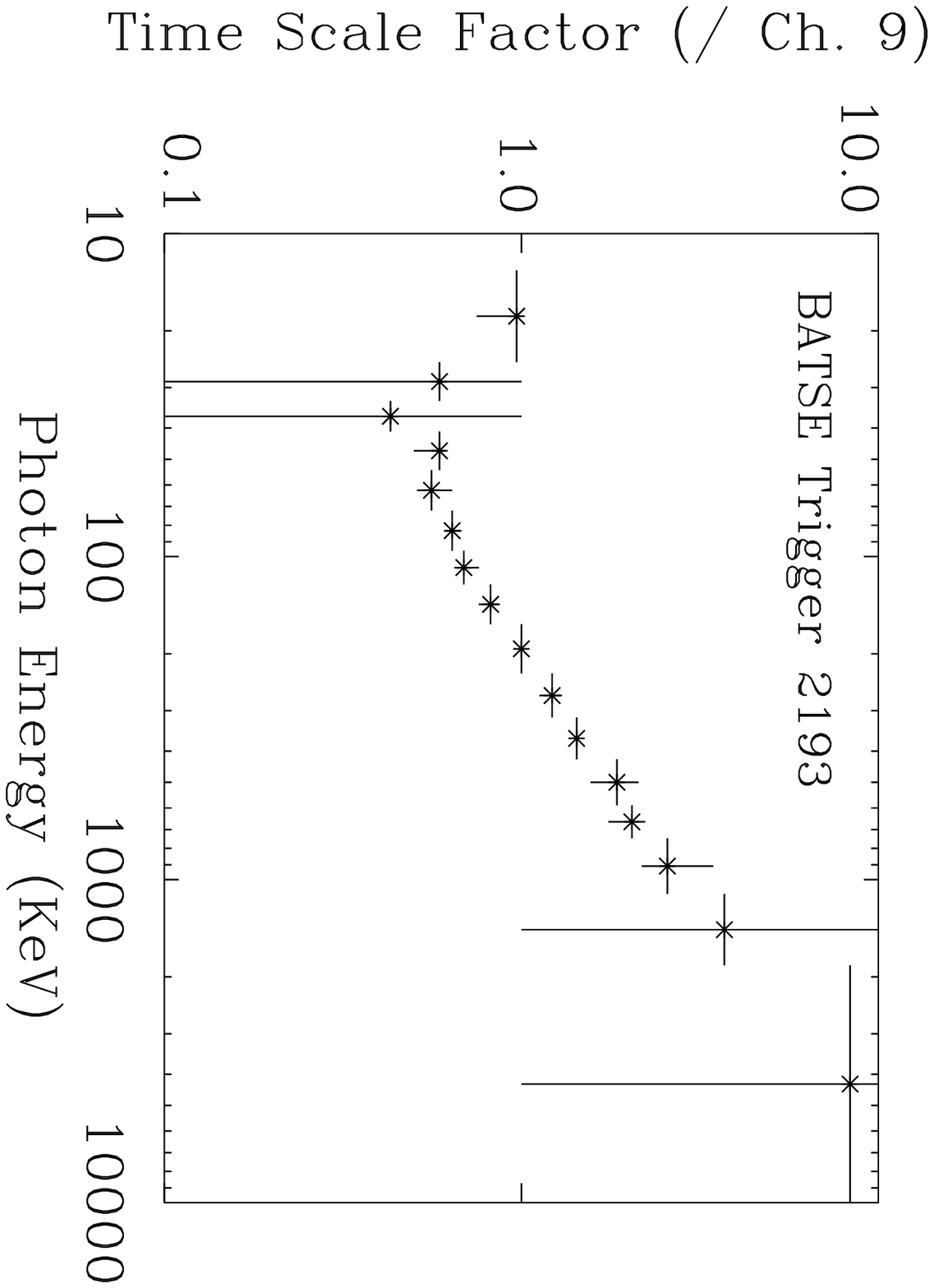]
{A plot of the best-fit temporal scale factor between each energy channel 
and channel 9, as a function of the energy midpoint of the photons in the energy channel.  Note that the log of the relative scale factors scale nearly linearly with the log of the photon energy of the channel.
\label{fig4}}

\figcaption[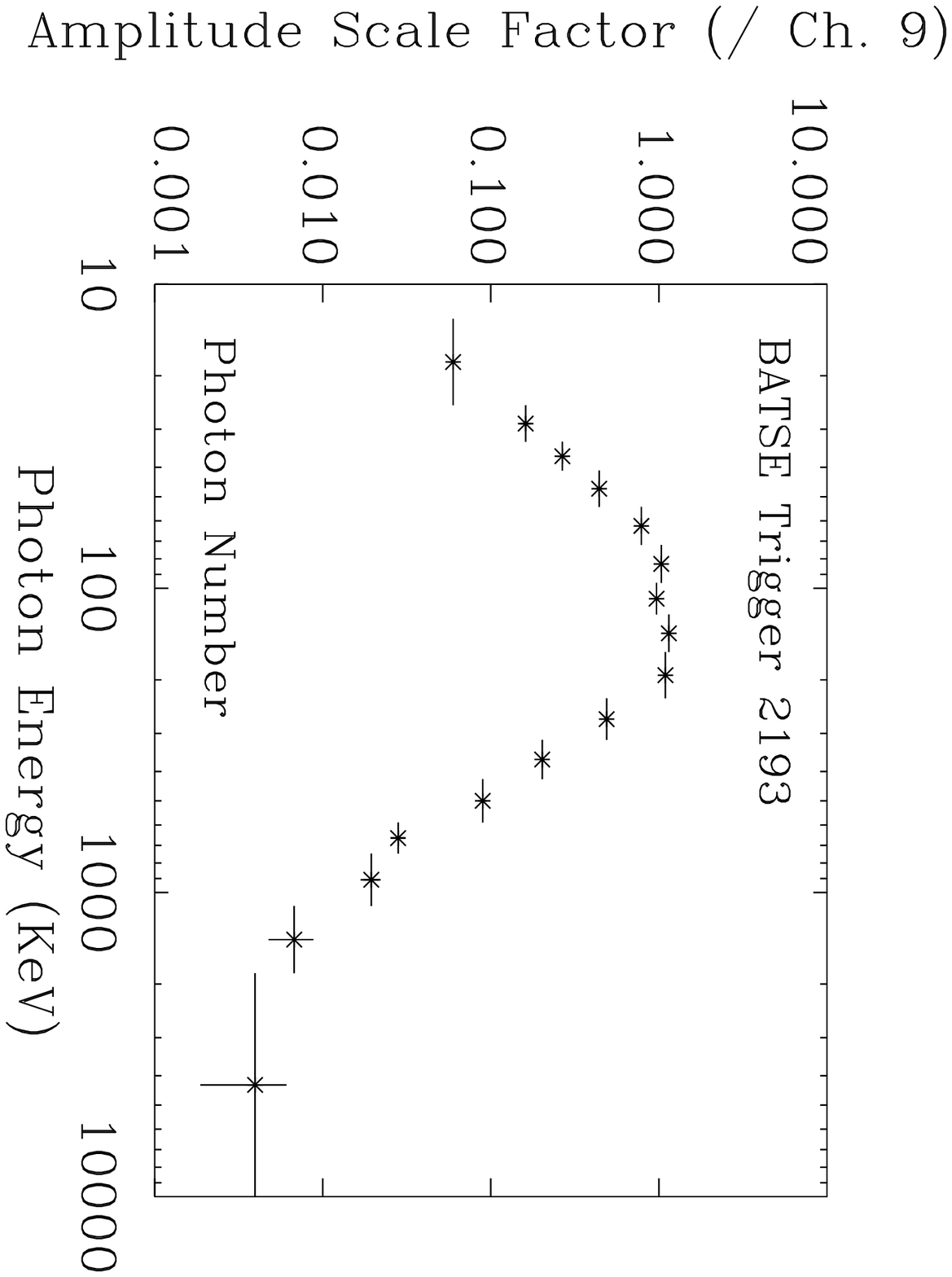]
{A plot of the best-fit amplitude scale factor between each energy channel 
and channel 9, as a function of the energy midpoint of the energy channel.
\label{fig5}}

\figcaption[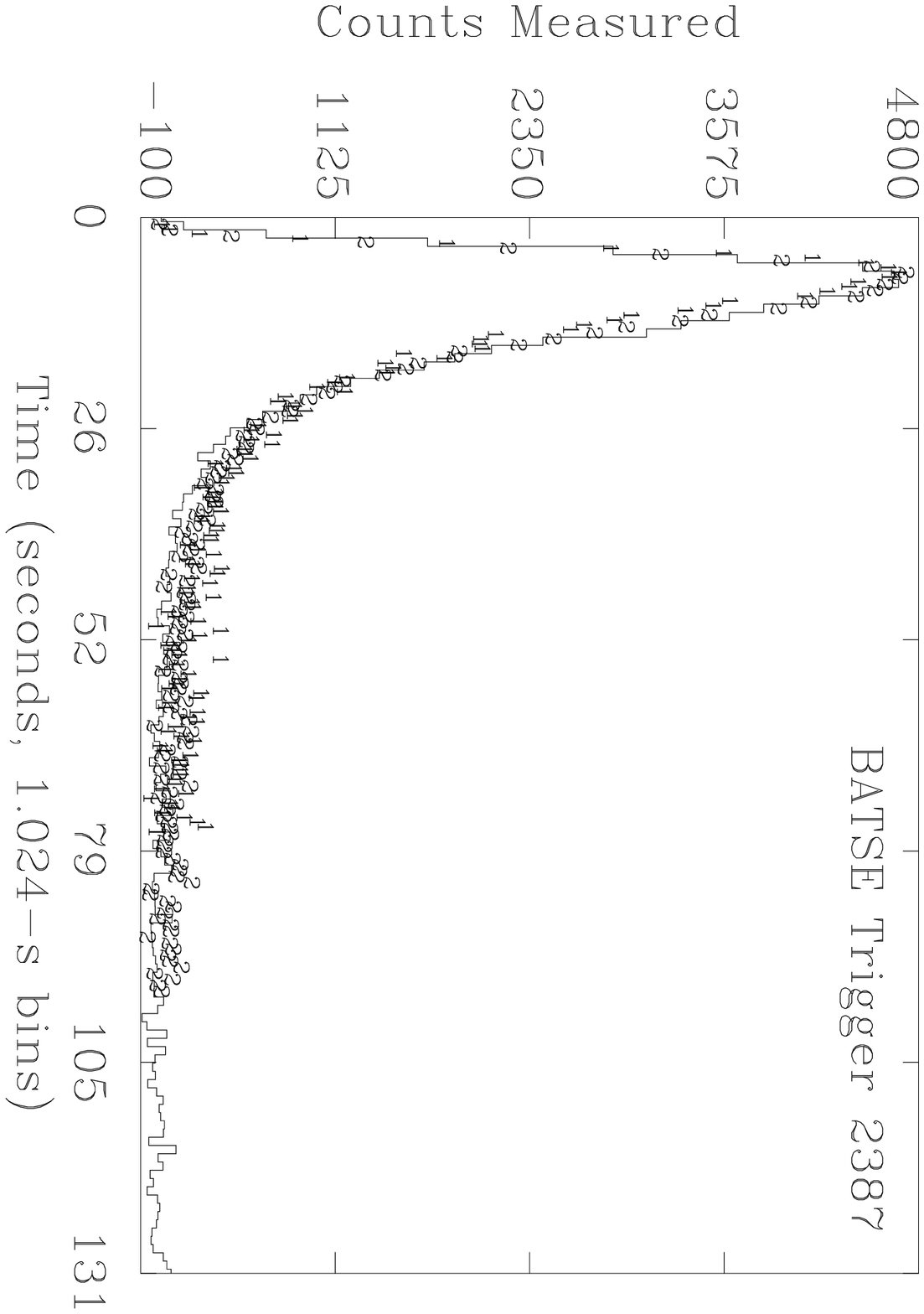]
{Light-curve plots of BATSE trigger 2387.  BATSE channel 3 is drawn in as a continuous line.  The scaled counts from energy channels 1 and 2 are plotted with the numbers ``1" and ``2" respectively.  Time has been set to zero at the beginning of each pulse.  Note how the scaled pulse in each energy channel has approximately the same shape.
\label{fig6}}

\figcaption[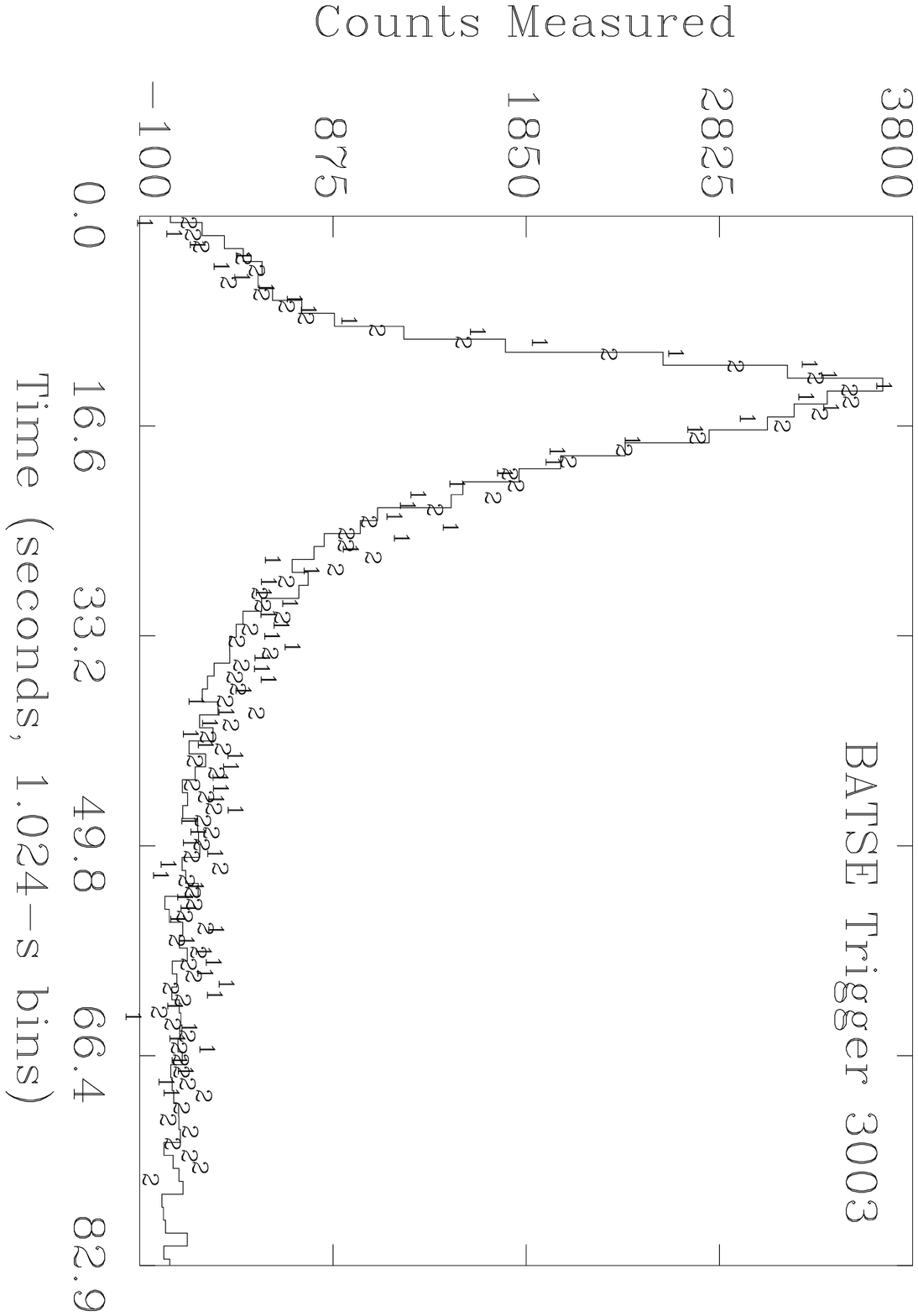]
{Light-curve plots of BATSE trigger 3003.  BATSE channel 3 is drawn in as a continuous line.  The scaled counts from energy channels 1 and 2 are plotted with the numbers ``1" and ``2" respectively.  Time has been set to zero at the beginning of each pulse.  Note how the scaled pulse in each energy channel has the same shape.
\label{fig7}}

\figcaption[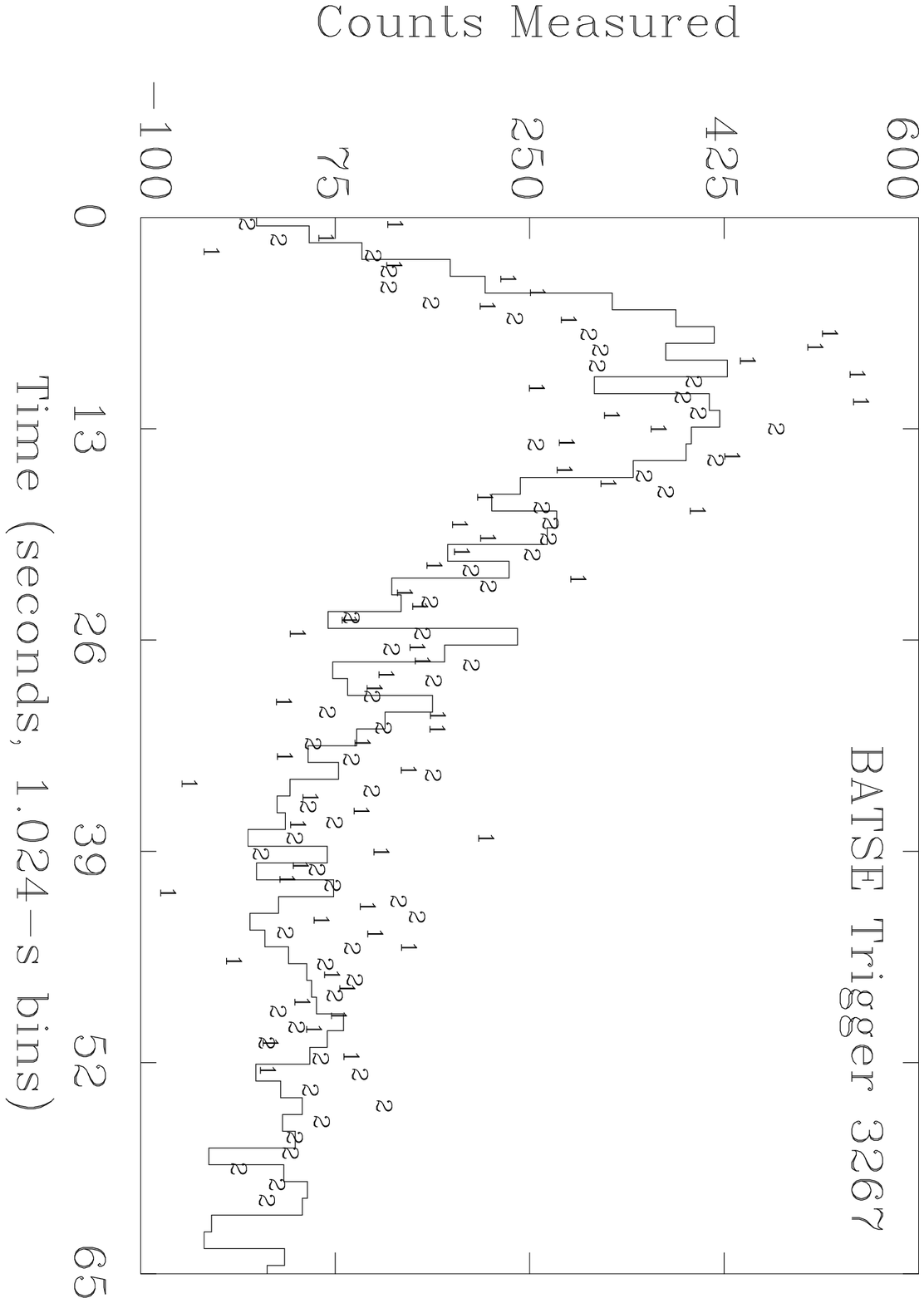]
{Light-curve plots of BATSE trigger 3267.  BATSE channel 3 is drawn in as a continuous line.  The scaled counts from energy channels 1 and 2 are plotted with the numbers ``1" and ``2" respectively.  Time has been set to zero at the beginning of each pulse.  Note how the scaled pulse in each energy channel has the same shape.
\label{fig8}}

\figcaption[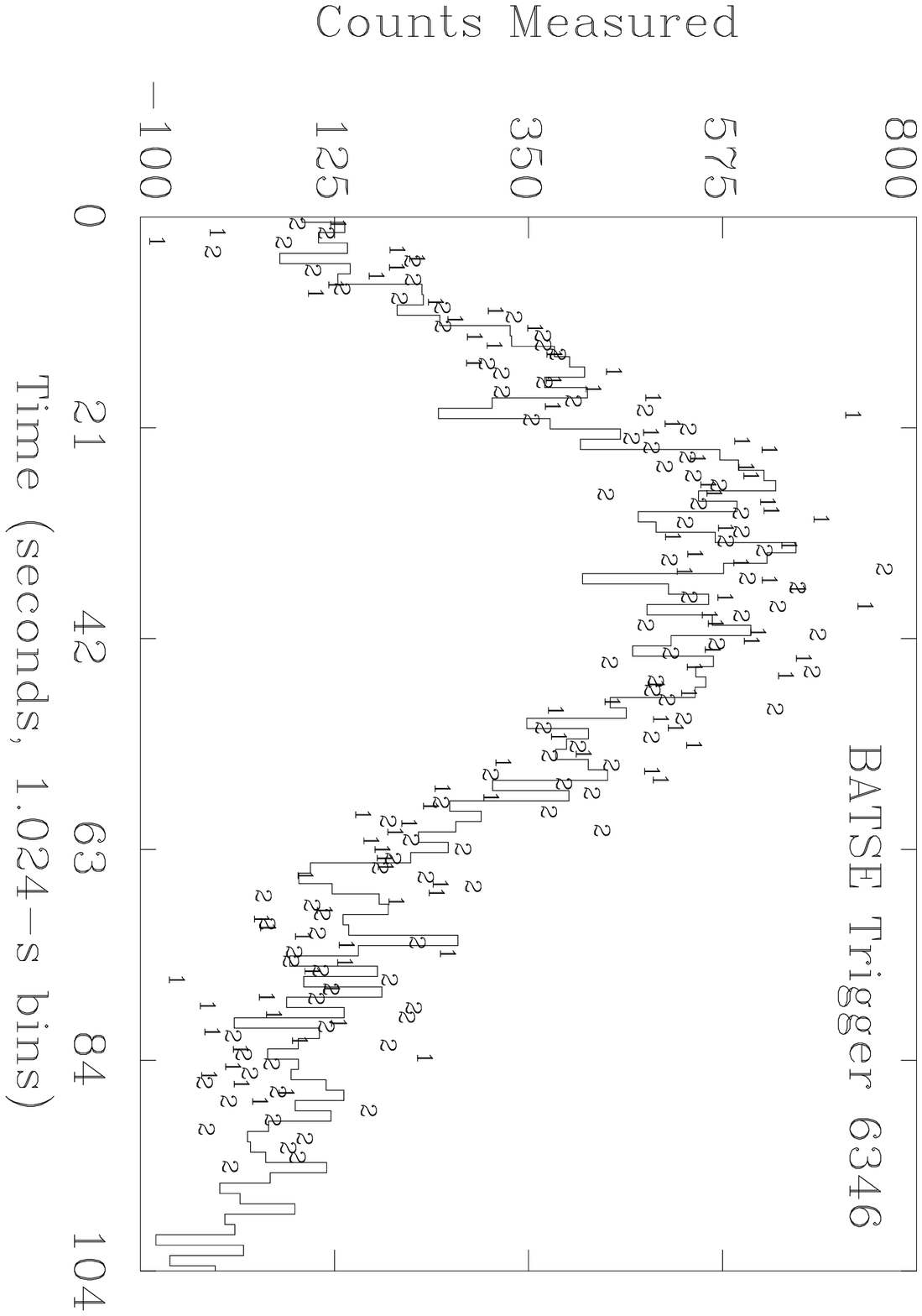]
{Light-curve plots of BATSE trigger 6346.  BATSE channel 3 is drawn in as a continuous line.  The scaled counts from energy channels 1 and 2 are plotted with the numbers ``1" and ``2" respectively.  Time has been set to zero at the beginning of each pulse.  Note how the scaled pulse in each energy channel has the same shape.
\label{fig9}}


\plotone{pulse2193_fig1.eps}

\plotone{pulse2193_fig2.eps}

\plotone{pulse2193_fig3.eps}

\plotone{pulse2193_fig4.eps}

\plotone{pulse2193_fig5.eps}

\plotone{pulse2193_fig6.eps}

\plotone{pulse2193_fig7.eps}

\plotone{pulse2193_fig8.eps}

\plotone{pulse2193_fig9.eps}

\end{document}